\documentclass[fleqn,10pt]{wlscirep}
\usepackage{xcolor}
\usepackage{graphicx}
\usepackage{amsmath}
\usepackage{amssymb}

\def\jasr{Adv. Space Res. }
\def\apj{Astrophys. J. }
\def\apjl{Astrophys. J. Lett.}

\def\aap{Astron. and Astrophys.}
\def\solphys{Sol. Phys.}

\def\mnras{Mon. Notices of RAS}
\def\nat{Nature}
\def\physrep{Phys. Rep.}

\def\ssr{Space Sci. Rev. }
\title{Oscillations of the baseline of solar magnetic field and solar irradiance on a millennial timescale}
\author[1,*]{Zharkova V.V.}
\author[2]{Shepherd S. J.}
\author[3]{Popova E.}

\affil[1]{Northumbria University, Department of Mathematics, Physics and Electrical Engineering, Newcastle upon Tyne, NE2 1XE, UK}
 \affil[2]{University of Bradford, School of Engineering, Bradford, BD7 1DP, UK}
\affil[3]{ Institute of the Physics of Earth, Moscow 123242; National Research University, Higher School of Economics, 101000, Moscow, Russia}
\affil[*]{valentina.zharkova@northumbria.ac.uk; valja46@gmail.com}

\begin{abstract}
Recently discovered long-term oscillations of the  solar background magnetic field associated with double dynamo waves generated in inner and outer layers of the Sun indicate that the solar activity is heading in the next three decades (2019-2055) to a Modern grand minimum similar to Maunder one. On the other hand, a reconstruction of solar total irradiance suggests that since the Maunder minimum  there is an increase in the cycle-averaged total solar irradiance (TSI) by a value of about $1-1.5$ $Wm^{-2}$ closely  correlated with an increase of the baseline (average) terrestrial temperature.  In order to understand these two opposite trends, we calculated the double dynamo summary curve of magnetic field variations backward one hundred thousand years allowing us to confirm strong oscillations of solar activity in regular (11 year) and recently reported grand (350-400 year) solar cycles caused by actions of the double solar dynamo.  In addition,  oscillations of the baseline (zero-line) of magnetic field  with a period of $1950\pm95$ years (a super-grand  cycle) are  discovered by applying a running averaging filter to suppress large-scale oscillations of 11 year cycles.   Latest minimum of  the baseline oscillations is found to coincide with the grand solar minimum (the Maunder minimum) occurred before the current super-grand cycle start. Since then the baseline magnitude became  slowly increasing towards  its maximum  at {\color{blue} $~$2700} to be followed  by its decrease and minimum at {\color{blue} $~$3700.} These oscillations of the baseline solar magnetic field are found associated with a long-term solar inertial  motion about the barycenter of the solar system and  closely linked to an increase of solar irradiance and terrestrial temperature in the past two centuries. This trend is anticipated to continue in the next five centuries that can lead to a further natural increase of the terrestrial temperature by 2.5$^\circ$C. 
\end{abstract}
\begin{document}

\flushbottom
\maketitle
%\thispagestyle{empty}

\section*{Introduction}
 
 Understanding of solar activity is tested by the accuracy of its prediction. The latter became very difficult to derive from the observed sunspot numbers and to fit sufficiently close into  the  prediction of a few future solar cycles or even into a single solar cycle until the cycle is well progressed\cite{pesnell08}. In many models  there is a onsistent disagreement for cycle 24 between the measured sunspot numbers and the predicted  ones. This disagreement  is a clear indication that there is something missing in the definition of solar activity by sunspot numbers, which are the collective product of many instruments and observers around the world. This discrepancy also indicates that the appearance of sunspots on the surface during a solar cycle is governed by the action of some physical processes of solar dynamo, which were not yet considered in the models. Moreover, it was first detected by Stix\cite{stix76} for cycle 22 and later confirmed by Zharkov et al.\cite{zhar08} for cycle 23 that the polarity of solar background magnetic field is always opposite to the leading polarity of sunspots while timing and locations of sunspot appearance on the solar surface are, in fact, governed by this  background magnetic field.

This research inspired us to investigate this solar background magnetic field s a new proxy of solar activity. In order to reduce dimensionality of any waves present  in the observational data of the background magnetic field, Zharkova et al.\cite{zharkova12} explored  a solar background magnetic field by applying Principal Component Analysis (PCA)  to the low-resolution full disk magnetograms captured in cycles 21-23 by the Wilcox Solar Observatory. This approach revealed not one but two principal components (PCs) with the  nearly-equal largest eigen values (strongest waves of solar magnetic oscillations) covering about 67$\%$ of the data by standard deviation\cite{shepherd14, zhar15}. The PCs are shown associated with two magnetic waves attributed to the poloidal  magnetic field \cite{popova13, zhar15} generated by a double dynamo with dipole magnetic sources in the inner and outer layers of the solar interior\cite{zhao2013}. These waves are found  to be asymmetric by default since they start in the opposite hemispheres while travelling with an increasing phase shift to the Northern hemisphere (in odd cycles) and  to the Southern hemisphere (in even cycles)\cite{zharkova12, zhar15}. The maximum  of solar activity for a given cycle (or double maximum for the double waves with a larger phase shift) occurs at the times when each of the waves approaches its maximum so that at the equal amplitudes the two waves can have resonant interaction. The hemisphere where it happens becomes the most active one, naturally accounting for the often-reported North-South asymmetry of solar activity in cycle 23\cite{zhar05,zhar08} and in a few other 11 year cycles\cite{temmer02, belucz_dikpati2013, shetye_dikpati2015}. 

 These two magnetic waves of poloidal field generated  during a solar cycle by the electromotive force in the two  layers (inner and outer) can be converted into  two waves of the toroidal magnetic field associated with sunspots \cite{Parker55, axel2005, jones10, popova13, popova17}.  The summary curves of the two waves of poloidal magnetic field produce the two magnetic waves of toroidal magnetic field and their summary curve,  whose modulus is closely associated with  solar activity defined by the averaged sunspot numbers\cite{shepherd14, zhar15}. The existence of two waves in the poloidal (and toroidal) magnetic fields generated in two layers, instead of a single one used in the most prediction models, and the presence of a variable  phase difference between the two waves can naturally explain the difficulties  in predicting the solar activity (or our summary curve) with a single dynamo wave\cite{2012ApJ...761L..13K}. 
 
 The two magnetic waves generated by magnetic dipoles  in two different layers of the solar interior generate with the electromotive force of solar dynamo\cite{Parker55} toroidal magnetic fields, or  magnetic loops which become sunspots on the surface. This interference is especially intense when the wave amplitudes become close, so that the waves can reach a resonance marking the maximum of solar activity for a given cycle. The hemisphere where these waves reach maxima, becomes the most active one. At the same time, the magnetic waves from the inner layer travel through the outer layer of solar interior to the solar surface and interfere with the magnetic waves generated in this outer layer.  At some  times the two waves, generated in inner and outer layers, appear to be in the anti-phase,  causing a disruptive interference.  This reduces dramatically the resulting wave magnitude and, thus, leads to significant reduction in production of toroidal fields, or sunspot numbers.  

The resulting summary curve, which is linked to the solar activity curve defined by the averaged sunspot numbers\cite{shepherd14}, restored backward for 3000 years shows about 9 grand cycles of 350-400 years, with the  times of their grand minima having remarkable resemblance to those reported from the  sunspot and terrestrial activity in the past millennia \cite{zhar17}: Maunder (grand) Minimum (1645-1715),  Wolf  grand minimum (1200),  Oort grand minimum (1010-1050),  Homer  grand minimum (800-900 BC), combined with the warming periods: medieval (900-1200), Roman  (400-10 BC) and other ones occurred between the grand minima. This approach allowed us to predict the modern grand solar minimum (GSM) approaching the Sun in 2020-2055\cite{zhar15}. This grand minimum  offers a unique opportunity for the space scientists and all people of the planet to witness in many details the modern grand minimum and to understand better the nature of solar activity.

Although, it was noted\cite{zhar17, zhar18} that Sporer minimum (1460-1550)  is not present in our summary curve, which instead during the same period of time shows  a standard grand cycle, the previous one to the modern grand cycle (17-21 centuries). Zharkova et al.\cite{zhar17, zhar18} reasonably argued  that  Sporer minimum is an artifact of the strongly increased at that time background radiation on the Earth  caused by the explosion of a very close (about 600-700 light years) supernova Vela Junior occurred in the southern sky. The radiation induced by this explosion  for this period has not been considered in the background radiation required  for the carbon dating method\cite{arnold_libby1949, baade1934} that could shift the dates by a few hundred years. 

These two-wave magnetic field variations were tested with Parker's two layer dynamo model with meridional circulation\cite{popova13, zhar15} showing  that the grand cycle variations of magnetic field are induced by a beating effect (with a period of 350-400 years) of the interference of  magnetic waves generated in each layer. These variations are affected by the changes of solar dynamo numbers in each layer describing a joint action  of solar differential rotation ($\Omega$-effect) and radial shear ($\alpha$-effect). It is assumed that the both dynamo waves are produced by dipole magnetic sources: one in the subsurface layer and the other deeply in the solar convection zone with the parameters in each layer to be rather different\cite{zhar15}. This difference  led to two magnetic waves, similar to those derived with PCA (see Fig.3 in Zharkova et al.\cite{zhar15}), which  travel from one hemisphere to another\cite{zhar15} with different but close frequencies and increasing phase shifts\cite{popova13} and producing the grand cycles of the similar durations and shapes as derived from the observations using PCA\cite{zharkova12}. 

The temporal and latitudinal patterns shown by two principal components defining dynamo waves generated in the inner and outer layers of the solar interior\cite{zhar15} can naturally account for the difference in observed magnetic fluxes in the opposite hemispheres reported by Shetye et al.\cite{shetye_dikpati2015}. Since the two double dynamo waves travel  with different phases, so that at a given moment they have different amplitudes (see Fig. 1 in Zharkova et al.\cite{zhar15}) that can explain a much larger  magnetic flux observed in the Northern hemisphere than in the Southern one at the descending phase of cycle 23 and a reversed  trend in the ascending phase of cycle 24  reported by Shetye et al.\cite{shetye_dikpati2015}.  In addition, these two waves generated in different layers gain close but not equal frequencies\cite{zhar15} because of different speeds of meridional circulation in each layer as suggested by Shetye et al.\cite{shetye_dikpati2015}, so that their interference naturally leads to a beating effect with the envelope oscillations (grand cycles) occurring at the frequency equal to a difference of frequencies of the individual waves\cite{zhar15}. The lengths of the individual grand cycles depend on a real time as shown in the summary curve of the solar activity extrapolated backward by two\cite{zhar15, popova17} and three millennia\cite{zhar17, zhar18} making some grand cycles shorter with higher amplitudes and the other ones longer with smaller amplitudes.  

 All these derivations of the observed magnetic waves, or principal components, generated by dipole magnetic sources were carried our purily from the solar magnetic data assuming that the Sun is an isolated system generating its own waves by its own (dynamo) rules. However,  Hays et al.\cite{hays76} shown that small planetary influences on the solar magnetism seen from the Earth can have long-term effects on the Earth's climate.  As established by Milankovich\cite{hays76, milankovich98} (see also $https://en.wikipedia.org/wiki/Milankovitch_-cycles$) there are various aspects of the Earth movements in the solar system, which can affect the terrestrial climate changes over many thousand years\cite{hays76, milankovich98}.  For example, the Earth axis tilt is shown to affect the terrestrial temperature variations with season and their durations, while the Earth orbit eccentricity and different type of precession define  long-term variations of the terrestrial temperature on a scale of 20, 40 and 100 thousand years as derived from the Antarctic glaciers\cite{rial03, akasofu10, abe13}. 

These orbital oscillations of the Earth rotation about the Sun strongly affect the solar irradiance and temperature on the Earth. Solar irradiance is accepted to be one of the important factors defining  the temperature variations on the Earth and other planets as it is the main source of the energy. Reconstruction of the cycle-averaged solar total irradiance back to 1610 suggests that since the end of the Maunder minimum there was the increase of the irradiance by a value of about $1–1.5$ $Wm^{-2}$\cite{krivova11}, or about 3$\%$ of the total solar irradiance. This increase is correlated rather closely with the oscillations of the terrestrial temperature baseline\cite{akasofu10}, which is found to steadily increasing since the Maunder minimum (e.g. recovering from the mini ice age).  Although, it is not clear yet if this trend in the terrestrial temperature and solar irradiance is caused directly by the increased solar activity itself or by some other factors of the solar-terrestrial interaction in the  whole solar system and human activities. 

\section*{Restoration of double dynamo waves for the past hundred millennia}
 As we shown earlier\cite{shepherd14, zhar15}, the resulting summary curve of the two magnetic waves detected with PCA can be used for prediction of  a solar activity usually associated with the averaged sunspot numbers\cite{shepherd14, zhar15}.  Let us explore  this summary curve\cite{zhar15}, as approximation of the solar activity on larger timescale of hundred millennia.

\begin{figure*}
\includegraphics[scale=0.5]{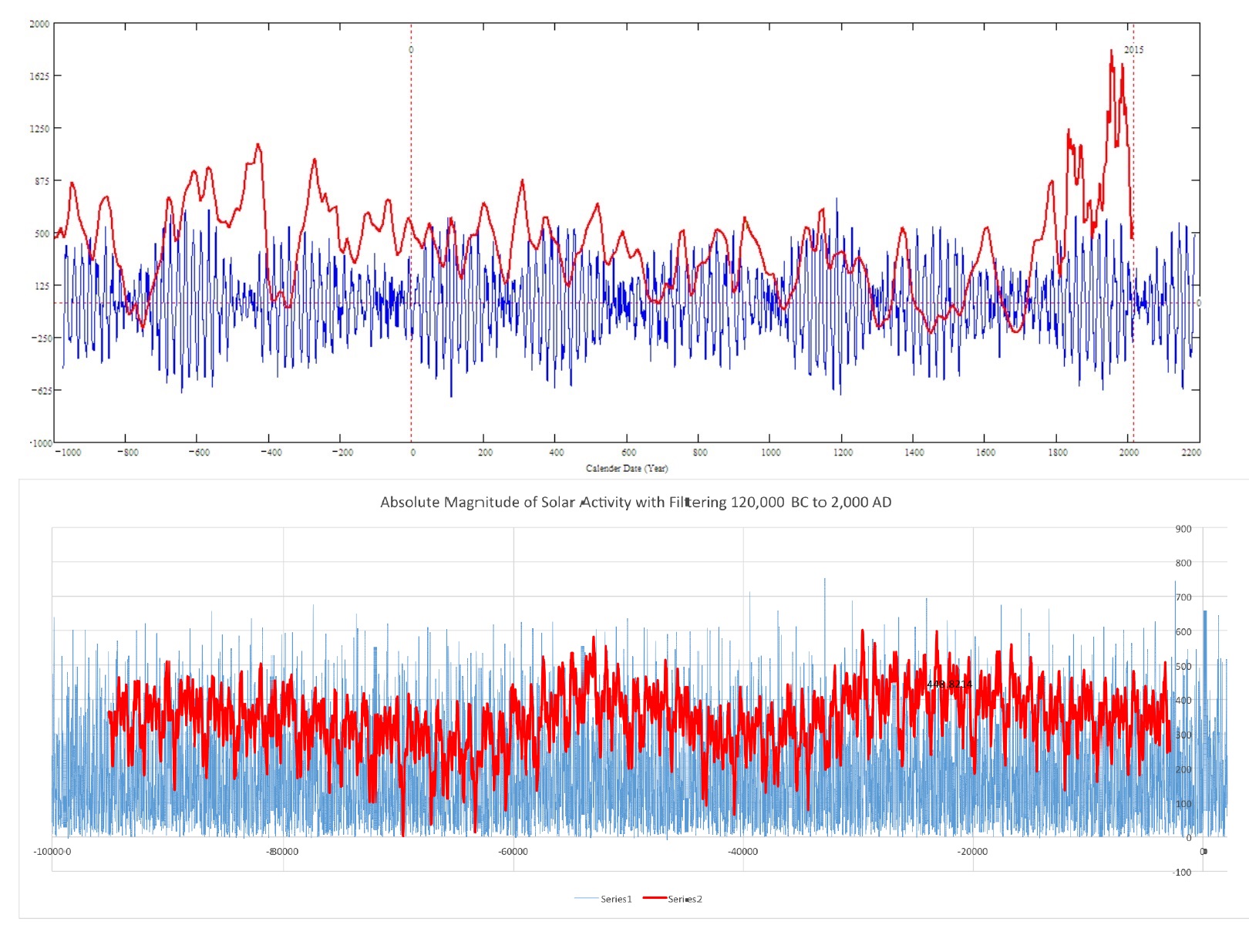}
  \caption{Top plot: solar activity prediction backwards 3000 years with a summary curve (blue line) of the two principal components (PCs) of solar background magnetic field (SBMF)\cite{zhar15} versus the reconstruction by Solanki et al.\cite{solanki04} (red line). The summary curve is derived from the full disk synoptic maps of Wilcox Solar Observatory for cycles 21-23, the reconstructed solar activity curve\cite{solanki04} was build by merging the sunspot activity curve (17-21 centuries) and a carbon-dating curve (before the 17 century). The bottom plot: the modulus summary  curve of two PCs associated with averaged sunspot numbers\cite{shepherd14,zhar15} calculated  backward 120 (one hundred twenty) thousand years. The red line shows the baseline oscillations with a period of about 40 thousand years likely associated with the Earth's axial tilt (obliquity)  (see the text for further details).} 
\label{summary100}
\end{figure*}

In Fig.\ref{summary100}  we  present 3000 years of the summary curve (top plot,  blue curve) calculated backward from the current date, on which we overplotted the graph of the {\color{blue} averaged sunspot numbers restored from the dendrochronologically dated radiocarbon concentrations derived by Solanki\ et al.\cite{solanki04}}(top plot, red curve). The solar irradiance curve prior 17 century was restored  from the carbon isotope $\Delta ^{14}$C abundances in the terrestrial biomass merged in 17 century till present days with the solar activity curve derived from the observed sunspot numbers. 

It can be noted that in many occasions the summary curve plotted backward for 3000 years in Fig. \ref{summary100}  reveals a remarkable resemblance to the sunspot and terrestrial activity reported for these 3000 years from the carbon isotope dating\cite{solanki04}. The summary curve shows accurately the recent grand minimum (Maunder Minimum) (1645-1715), the other grand minima: Wolf  minimum (1300-1350), Oort minimum (1000-1050), Homer minimum (800-900 BC); also the Medieval Warm Period (900-1200), the Roman Warm Period (400-150 BC) and so on.  These  grand minima and grand maxima reveal the presence of a grand cycle of solar activity with a duration of about 350-400 years that is similar to the short term  cycles detected in the Antarctic ice \cite{rial03, akasofu10}.  The 11/22 and 370-400 year cycles were also  confirmed in other planets by the spectral analysis of solar and planetary oscillations \cite{scafetta2014, obridko2014}. The next Modern grand  minimum of solar activity is upon us in 2020-2055 \cite{zhar15}.  

Zharkova et al.\cite{zhar15} pointed out that longer grand cycles have a larger number of regular 11 year cycles inside the envelope of a grand cycle but their amplitudes are lower than in shorter grand cycles. This means that  there are significant modulations of the magnetic wave frequencies generated for different grand cycles in these two layers: a deeper layer close to the bottom of the Solar Convective Zone (SCZ) and shallow layer close to the solar surface whose physical conditions derive the dynamo wave frequencies and amplitudes. The larger the difference between these frequencies the smaller the number of regular 22 years cycles inside the grand cycle and the higher their amplitudes.   Later Popova et al.\cite{popova17} have also shown that the reduced solar activity during Dalton minimum (1790$-$1820), which was weakly present in the summary curve for dipole sources \cite{zhar15}, is reproduced much closer to the observations of averaged sunspot activity by consideration of the quadruple components of magnetic waves, the next two eigen vectors obtained with PCA\cite{zharkova12}, produced by quadruple magnetic sources. 

In addition, in Fig.\ref{summary100} (bottom plot) we present the summary curve simulated for 100 000 years backwards from now (blue line), on which we over-plotted the averaged baseline curve (red curve) filtering large cycle oscillations with a running averaging filter of 25 thousand years. This plot reveals the baseline oscillations of about 40,000 (forty thousand)years (see the periodic function appearing between 20K and 60K years in the bottom plot), which are likely to be the oscillations caused by the Earth axis tilt (obliquity)\cite{hays76, milankovich98}, or the variations of the Earth axis tilt between $22.1^\circ$ and $24.5^\circ$ (the current tilt is $23.44^\circ$). This Earth obliquity effect is incorporated into the summary curve derived by us from the solar magnetic observations. This indicates that the measurements of a magnetic field of the Sun from the Earth, or from the satellites on the orbit close to Earth, contain also the orbital effects of the Earth rotation about the Sun and of any other motion by the Sun itself, which we intend to explore further in the sections below.

\subsection*{Detection of the baseline oscillations of solar magnetic field} \label{osc2000}
\begin{figure*}
\includegraphics[scale=0.53]{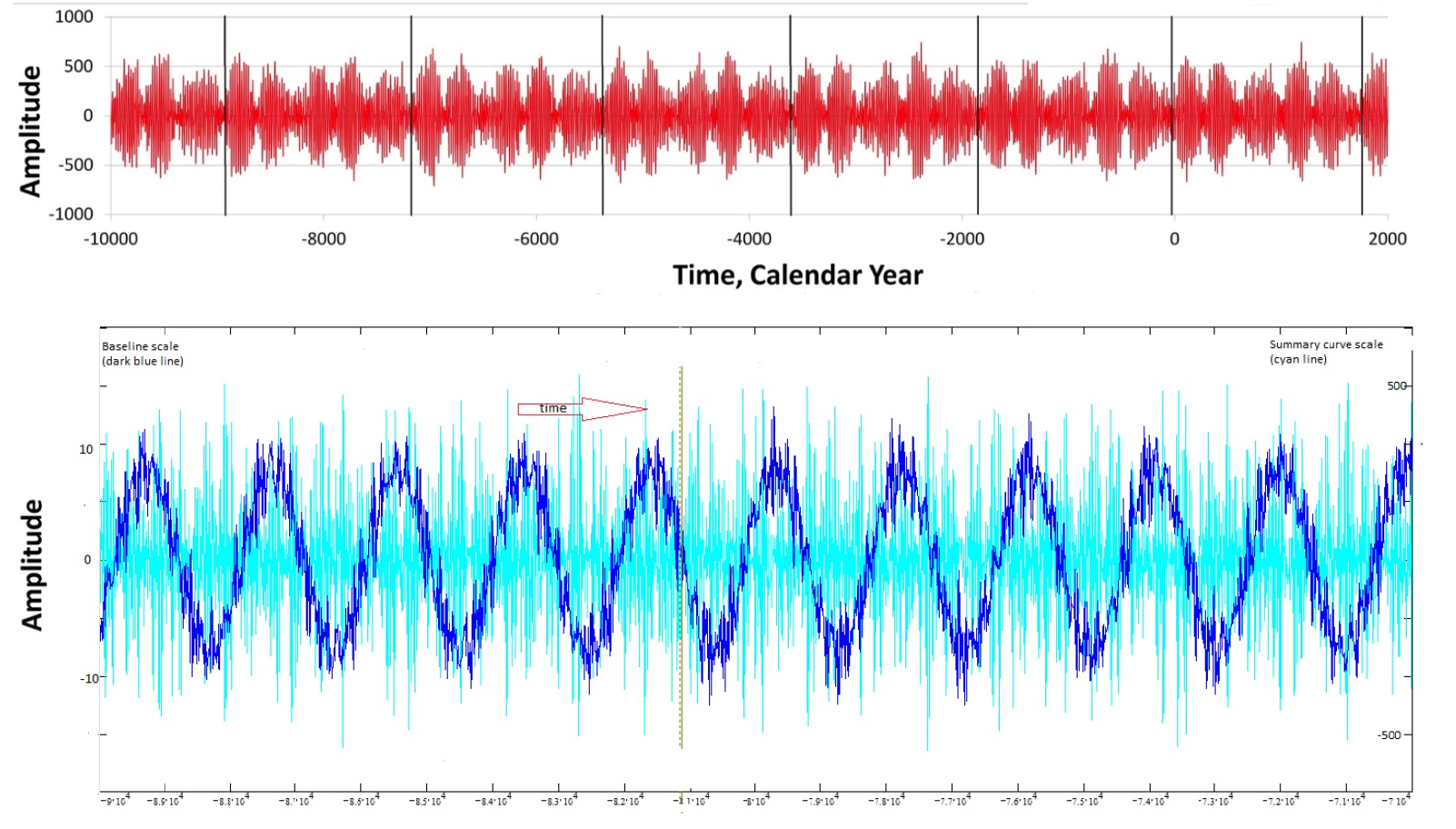}
\caption{ Top plot:  the summary  curve of two magnetic field waves, or PCs, calculated backward ten thousand years from the current time. The vertical lines define the similar patterns in five grand cycles repeated every 2000-2100 years (a super-grand cycle). Bottom plot:  the oscillations of the summary curve (cyan line) calculated backward from 70K to 90K years overplotted by the oscillations of a magnetic field baseline, or its zero line (dark blue line) with a period of about 1950$\pm$95 years. The baseline oscillations are obtained with averaging running filter of 1000 years from the summary curve suppressing large scale cycle oscillations.   The left  Y-axis shows the scale of variations of the baseline magnetic field, while the right  Y-axis presents  the scale of variations of the summary curve.  } 
\label{periods}
\end{figure*}

In Fig.\ref{periods} we present 20 000 years of the summary curve (between 70 and 90 thousand years backward), in which, in addition to the grand cycles of $~$350-400 years of solar activity, there are also indications marked by the vertical lines separating larger super-grand cycles (the top plot).   By comparing  in Fig.\ref{periods} (top plot) the semi-similar features (between the vertical lines) of the repeated five grand cycles of total with a duration of about 2000-2100 years, one can see a striking similarity of the shapes of these 5 grand cycles, which are repeated about 9 (nine) times during the 20 000 years. 

In order to understand the nature of these super-grand oscillations and to derive the exact frequency/period of this super-grand cycle, let us filter out large  oscillations of 11/22 year solar cycles with the running averaging filter (1000 years). The resulting baseline oscillations are shown by a dark blue curve in Fig.\ref{periods} (bottom plot) over-plotted on the summary curve (light blue curve) taken from the summary curve calculated backwards between 90 and 70 thousand years. For a comparison, the left Y-axis gives the range of variations of the baseline curve (-10,10) while the right side Y-axis produces the same for the summary curve (-500,500). The baseline variations are, in fact, the variations of the zero-line of the summary curve, which are too small to observe on this curve without filtering large-scale oscillations of 11 year cycles.
 
It is evident that the dark blue line in Fig.\ref{periods} (bottom plot) shows much (15 times) smaller oscillations of the baseline of magnetic field with a period of $T_{SG}=1950 \pm 95$ years, which is incorporated into the magnetic field measurements of the summary curve (light blue curve). The baseline oscillations show a very stable period  occurring during the whole duration of simulations of 120 thousand years, for which  the summary curve was calculated. This means  that this oscillation of the baseline magnetic field on a millennial timescale has to be induced by a rather stable process either inside or outside the Sun.  This baseline oscillation period is very close to the 2100-2400  year period {\color{blue} called Hallstatt's cycle} reported from the other observations of the Sun and planets\cite{obridko2014, Steinhilber12, scafetta2014, Fairbridge1987}. 
   
To understand the nature of these oscillations, we decided to compare these oscillations with the sunspot numbers restored from the carbon isotope abundances  for the past 10000 years {\color{blue} by Vieira at al. \cite{vieira2011}} as presented for 3000 years in Fig. \ref{summary100} (top plot, red curve). In Fig.\ref{irradiance} the irradiance curve by {\color{blue}Vierra et al.}\cite{vieira2011} was plotted for the current and the past grand cycles as follows: the summary curve of magnetic field (light blue line), the oscillations of the baseline (dark blue line) and the restored solar irradiance\cite{vieira2011} (magenta line), which was slightly reduced in magnitude in the years 0-1600, in order not to obscure the baseline oscillations. The dark rectangle indicates the position of Maunder Minimum (MM) coinciding with the minimum of the current baseline curve and the minimum of solar irradiance.  After the MM the baseline curve is shown growing for the next 1000 years (e.g. until 2700). During the current cycle years the solar irradiance curve\cite{vieira2011} follows this growth of the baseline (with {\color{blue}Spearman's} correlation coefficient about 0.68). 

For the further information we present in Fig.\ref{irradiance} (bottom plot) the variations of the Earth temperature for the past 140 years as derived by Akasofu\cite{akasofu10} with the solid dark line showing the baseline increase of the temperature, blue and red areas show natural oscillations of this temperature caused by combined terrestrial causes and solar activity. The trend derived by Akasofu\cite{akasofu10} shows the increase of the terrestrial temperature by 0.5$^\circ$C per 100 years. This temperature growth  is also expected to continue in the next 700 years until $~$2700. Although, if it follows the baseline curve, this growth could  not be linear  as it was at the early years shown in the Akasofu's curve\cite{akasofu10} but will have some saturation closer to the maximum as any periodic functions (sine or cosine) normally have.  

\subsection*{Links of the baseline oscillations with solar inertial motion (SIM)}  

Principal components  and their summary curve were detected from the solar background magnetic field oscillations produced  in the Sun. Large part of these oscillations related to 11 year solar cycle and 350-400 year grand cycle are well accounted for by the solar dynamo waves generated  dipole magnetic sources in inner and outer layers\cite{zhar15}. They can explain the magnetic field oscillations with a grand cycle by the beating effects of the two waves generated these two layers.  However, it  is rather difficult to find any mechanism in the solar interior that can explain much weaker and longer oscillations of the baseline of magnetic field. Therefore, we need to look for some external reasons for these oscillations.  

Kuklin\cite{Kuklin1976} first suggested that solar activity on a longer timescale can be affected by the motion of  large planets of the solar system. This suggestion was later developed by Fairbridge\cite{Fairbridge1987}, Charvatova \cite{charvatova1988} , Shirley\cite{Shirley1990} and Palus\cite{palus2007}  who found that the Sun, as a central star of the solar system, is a subject to the inertial motion around the barycenter of the solar system induced by the motions of the other planets (mostly large planets, e.g. Neptune, Jupiter and Saturn). 

Solar inertial motion (SIM) is the motion of the Sun around this barycenter of the solar system as shown in Fig.\ref{sim_cone} reproduced from the paper by Richard Mackey\cite{mackey07}]. Shown here are three complete orbits of the Sun, each of which takes about 179 years. Each solar orbit consists of about eight, 22-year solar cycles\cite{mackey07}. The total time span shown in Fig. \ref{sim_cone} is, therefore, three 179-year solar cycles\cite{Fairbridge1987}, for about 600-700 years.  The Sun rotates around the solar system barycenter inside the circle with a diameter of about $\Delta=4.3R_{Sun}$, or $\Delta$=$4.3\times 6.95\cdot10^5$ =$2.9885\cdot10^6$ km, where $R_{Sun}$ is a solar radius. This schematic drawing illustrates sudden shifts in the solar inertial motion (SIM) as the Sun travels in an epitrochiod-shaped orbit about the center-of mass of the solar system. 

 The SIM has very complex orbits induced the trifall positions of large planets achieved  for different planet configurations changing approximately within 370 years as indicated by Charvatova\cite{charvatova1988}. She also claimed that there is a a larger period of 2200-2400 years related to the full cycle of the planet positions in their rotation around the Sun\cite{charvatova2000} (see Fig. \ref{sim1}  from Charvatova's paper).  Since the SIM occurs for the Sun observed from the Earth, we believe,  only the SIM  can define the weak oscillations of the baseline of solar magnetic field reported above. 

Although, unlike Fairbridge\cite{Fairbridge1987} and Charvatova\cite{charvatova1988}, we do not propose a  replacement of the solar dynamo role in solar activity with the effects of large planets, or solar inertial motion. This replacement would be very unrealistic from the energy consideration\cite{zaqarashvili97}  because the tidal effects of the planets are unable to cause a direct effect on the dynamo wave  generation in the bottom of solar convective zone (SCZ).  

However, in the light of newly discovered double dynamo effects in the solar interior\cite{zhar15} the planets can surely perturb properties of the solar interior governing the solar dynamo in the outer layer, such as solar differential rotation, or $\Omega$-effect, governing migration of a magnetic flux  through the outer layer to its surface, and those of $\alpha$ effect, that can change the velocity of meridional circulation.  This leads to the dynamo waves in this outer layer with the frequency slightly different from that than in the inner layer, and, thus to the beating effects caused by interference of these two waves and to grand cycles discussed above\cite{zhar15}.

Although, Abreu et al (2012)\cite{Abreu2012} suggested that the tidal forces of large planets can excite gravity waves at the tachocline, which can propagate to the surface balanced by buoyancy of the solar interior\cite{Goldreich89, Barker10} and insert a net tidal torque in the small region between the tachocline and radiative zone. At the same time the shape of tachocline was inferred from helioseismic observations with prolate  geometry to show the ellipticity 1000 times higher than at the photospheric level\cite{Charbonneau1999}. 

Using this finding, Abreu et al.\cite{Abreu2012} suggested that a possible planetary torque can appear from the non-spherical tachocline  and modulate the dynamo waves properties generated there (Abreu et al, 2012). The authors used either $^{10}Be$ or $^{14}C$ isotope production rates of the terrestrial proxies to derive the various periods of solar and terrestrial activity using wavelet analysis and found the periods close to 370 and 2200 years reported above.  Although, the periods of this activity found by Abreu et al.\cite{Abreu2012}  were later objected by Cameron and Shussler\cite{Cameron13}, who argued that these activity periods are random and do not have a real causal force. This dialog demonstrated that the absence of long-term solar data was the obstruction for the accurate detection of shorter periods of  the solar-terrestrial activity.
 
However, a detection with PCA of the solar background magnetic field\cite{zharkova12,zhar15} and the HMI helioseismic observations\cite{zhao2013} of two layers in the solar interior with different directions and speeds of meridional circulation where  two dynamo waves can be generated either by dipole \cite{zhar15} or dipole plus quadruple\cite{popova17} magnetic sources  lifts these rather rigid requirements for the planetary torque to act very deeply inside the Sun at its tachocline. Instead, the planetary torque can affect differently the buoyancy and differential rotation of the convective zone in the  outer layers in both hemispheres, thus, producing there rather different $\alpha$- and $\Omega$-effects and different velocities of meridional circulation compared to those in the inner layer near the bottom of the tachocline. 

These parameters, in turn, are likely to be the effective contributors governing the frequencies and  phases of the dynamo waves  in the outer layer, thus, producing  the resulting beating frequencies obtained from the summary wave caused by these two wave interference. From a description of Milankovich cycles, we can assume that the larger period (40K years)  of oscillations of the baseline shown in Fig.\ref{summary100} is likely caused by  precession of the tile of the axis of  Earth's  rotation relative to the fixed stars\cite{milankovich98}.

\subsection*{Effects of SIM on solar irradiance}
JPL ephemeris  produce the variations of Sun-Earth distance  affected by four large planets (Jupiter, Saturn, Neptune and Uranus)\cite{folkner2014}. The S-E distance is found, in average, linearly reducing  from 1700 until 2700 with the rate of 0.00027 au per 100 years, or 0.0027 au per 1000 years\cite{folkner2014}.  It can be shown following the squared inverse distance law, the solar irradiance caused by the change of Sun-Earth distance caused by SIM can be increased by more then 0.5$\%$ in the closest point to the Earth and decreased by the same amount in the most distant point. This results in the increase of solar irradiance from 1700 until 2019 by about 2.0 W$\cdot m^{-2}$ that is close to 1.5 W$\cdot m^{-2}$  reported in 2011 by Krivova et al.\cite{krivova11}. The occurrence of these closest and most distant points vary in time with a period of about 2100-2400 years \cite{Steinhilber12, charvatova2000}.  In addition, there are much smaller variations of the distances imposed by non-circular SIM orbits where the Sun is found  unevenly shifted either to the Earth's aphelion (summer solstice) or to its perihelion (winter solstice). 

In order to understand how this SIM motion  would affect the solar irradiance at the Earth orbit {\color{blue} on a large millennial scale}, let us look at the drawing of the Earth motion around the Sun (Fig.\ref{earth_orbit}). If the  Sun was stationary and located in the focus of the Earth orbit, then the solar irradiance and, thus, the seasons on the Earth are defined by the position of our planet on the orbit around the Sun. In the aphelion   ($1.521\times10^{8}$ km from the focus where the Sun is located, 21-24 June, position 1) there is a summer in the Northern hemisphere and winter in the Southern hemisphere. While in the perihelion ($1.47\times10^{8}$ km from the Earth orbit focus, 21-24 December, position 2) there is a summer in the Southern and winter in the Northern hemispheres. The seasons are caused by the increase or reduction of the solar irradiance caused, in turn, by the inclination of the Earth's axis towards or from the Sun.

 Since the Sun moves around the solar system barycenter {\color{blue} partially dragging the Earth with it}, it implies that it {\color{blue} still slightly shifts} around the main focus of the ellipse orbit  being either closer to its perihelion or to its aphelion. If the Earth rotates around the Sun undisturbed by inertial motion, then the distances to its perihelion will be {\color{blue}$1.471\times10^{8}$} km and to it aphelion {\color{blue} $1.521\times10^{8}$} km. The solar inertial motion means for the Earth that the distance between the Sun and the Earth has to significantly change {\color{blue} (up to 0.0027 of a.u, or 402395 km by 2700)} at the extreme positions of SIM, and so does  the average solar irradiance, which is inversely proportional to the squared distance between the Sun and Earth. 
 
If during SIM {\color{blue} phases} the Sun moves closer to perihelion and the spring equinox (positions 2), the distance between the Sun and Earth will be the shortest at perihelion approaching about {\color{blue} $1.46726\cdot 10^{8}$}km while at aphelion it will increase to {\color{blue} $1.52474\cdot 10^{8}$} km.  This means at these times the Earth  would receive higher than usual solar irradiance (that can lead to higher terrestrial temperatures\cite{Eddy1976, Shirley1990, akasofu10}), while approaching its perihelion during its winter and spring (warmer winters and springs in the Northern hemisphere and summers and autumns in the Southern one). At the same time, when the Earth moves to its aphelion, the distance between the Earth and Sun is increased because of SIM resulting in the reduced solar irradiance  during summer and autumn in Northern and winter/spring in the Southern hemispheres. This scenario with the increase of solar irradiance and terrestrial temperature was likely to happen during the millennium prior the Maunder Minimum.
 
If the Sun moves in its SIM closer to Earth's aphelion (position 1) and autumn equinox as it is happening in the current millennium starting from Maunder Minimum, then the distance between Sun and Earth will be shorter at the aphelion approaching {\color{blue} $1.51726\times10^{8}$} km at aphelion, or during the summer in the Northern and winter in the Southern hemispheres, and longer at the perihelion approaching {\color{blue} $1.47474\times10^{8}$ } km, or during a winter in the Northern and summer in the Southern hemispheres.  Hence, at this SIM position of the Sun, the Earth in aphelion should receive higher solar irradiance (and  temperature\cite{Eddy1976, Eddy1983, Shirley1990}) during the Northern hemisphere summers and Southern hemisphere winters.  When the Earth moves to its perihelion, the distance to the Sun and thus, the solar irradiance will become lower leading to colder winters in the Northern hemisphere and colder summers in the Southern one. This is what happening in the terrestrial temperature in the current millennium starting since Maunder minimum until $\approx$2700.
 
 Hence, it is evident that the oscillations of the solar inertial motion around the barycentre of the solar system should  produce the very different variations of  solar irradiance in each hemisphere of the Earth at different seasons.  These variations occur in addition to any other variations of the solar irradiance caused by larger variations of the solar activity itself caused by the action of  solar dynamo.  Currently, the solar system is at the SIM phase when the Sun moves towards the aphelion (position 1). This  is expected to lead to a steady increase for another 700 years of the baseline magnetic field. 
 
\begin{figure*}
\includegraphics[scale=0.55]{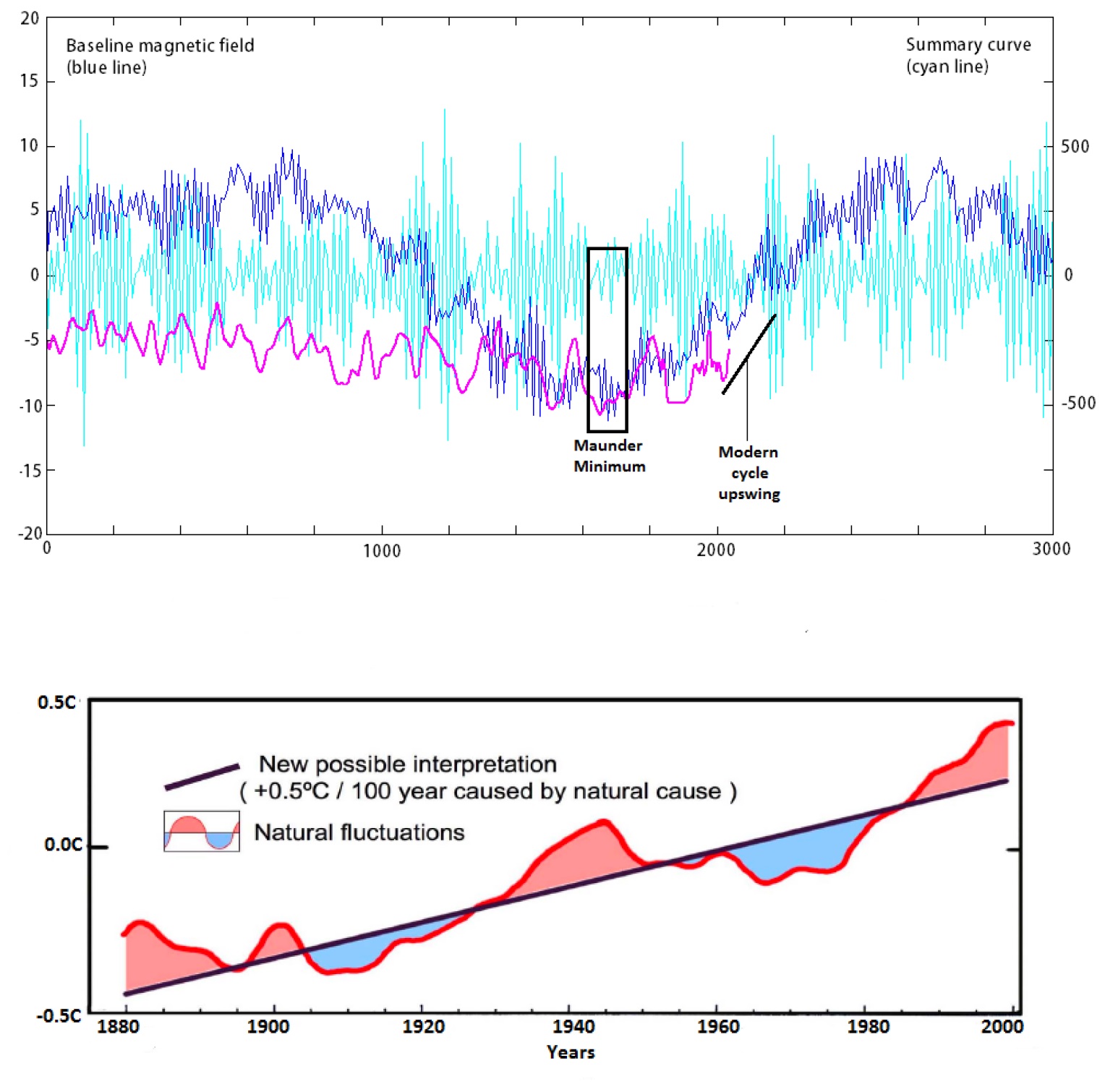}
\caption{ Top plot: the close-up view of the oscillations  of the baseline magnetic field (dark blue curve) in the current and past millennia with a minimum occurring during Maunder Minimum (MM). The irradiance curve (magenta line) presented from Krivova and Solanki\cite{krivova11, solanki11}  overplotted on the summary curve of magnetic field (light blue curve\cite{zhar15}). Note the irradiance curve is slightly reduced in magnitude in the years 0-1400 to avoid messy curves. The dark rectangle indicates the position of MM coinciding with the minimum of the current baseline curve and the minimum of the solar irradiance\cite{krivova11, solanki11}. The scale of the baseline variations are shown on the left hand side of Y axis, the scale of the summary curve - on the right hand side. Bottom plot: variations of the Earth temperature for the past 140 years derived by Akasofu\cite{akasofu10} with the solid dark line showing the baseline increase of the temperature, blue and red areas show natural oscillations of this temperature caused by combined terrestrial causes and solar activity. The increase of terrestrial temperature is defined by 0.5$^\circ$C per 100 years\cite{akasofu10}. } 
\label{irradiance}
\end{figure*}

\begin{figure*}
 \includegraphics[scale=0.5]{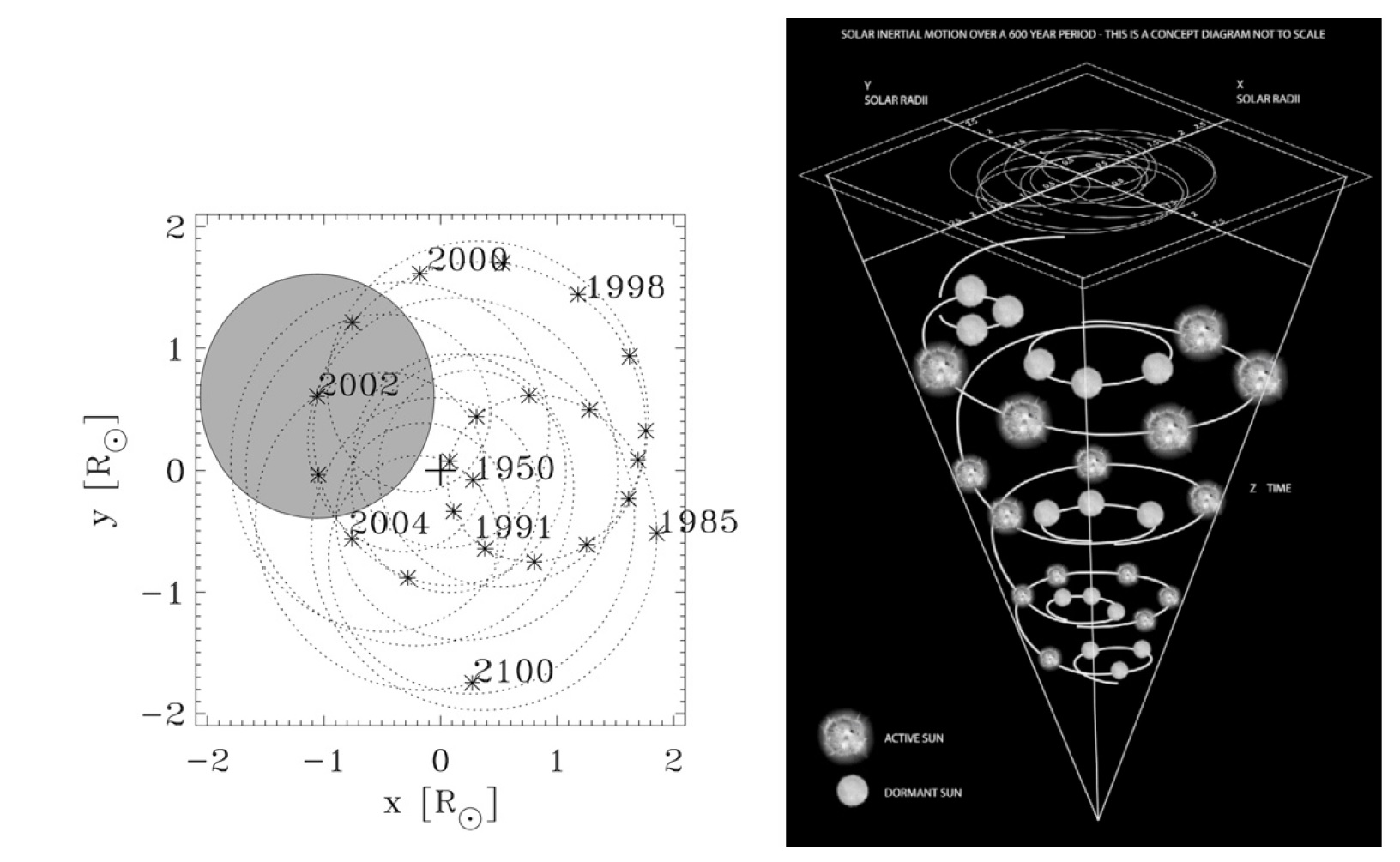}
\caption{Left plot: the example of SIM trajectories of the Sun  about the barycenter calculated from 1950 until 2100\cite{palus2007}. Right plot: the cone of expanding SIM orbits of the Sun\cite{mackey07} with the top  showing 2D orbit projections similar to the left plot. Here there are three complete SIM orbits of the Sun, each of which takes about 179 years. Each solar orbit consists of about eight, 22-year solar cycles\cite{mackey07}. The total time span is, therefore, three 179-year solar cycles\cite{Fairbridge1987}, or about 600 years. Source: Adapted from Mackey (2007)\cite{mackey07}. Reproduced with permission from the Coastal Education and Research Foundation, Inc.} 
\label{sim_cone}
\end{figure*}

\begin{figure*}
\includegraphics[scale=0.58]{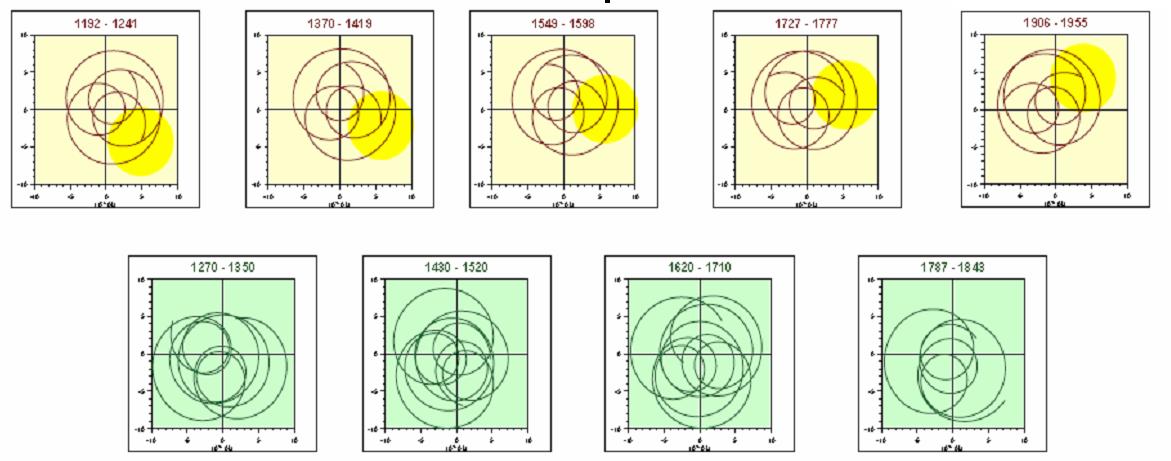}
\caption{ Schematic presentation of the solar inertial motion (SIM) about the barycenter of the solar system defined by the gravitational forces of large planets in the plane of ecliptics (the top plane in Fig.\ref{sim_cone}) for different time intervals shown in the top of each sub-figure (reproduced from Charvatova\cite{charvatova2000}.  The location of the Sun at the end of the period is shown by the yellow circles. Top row represent the ordered SIM affected by symmetric positions of large planets with respect to the Sun, while the bottom row shows the disorganized SIM with more random positions of large planets. } 
\label{sim1}
\end{figure*}

\begin{figure*}
\includegraphics[scale=0.31]{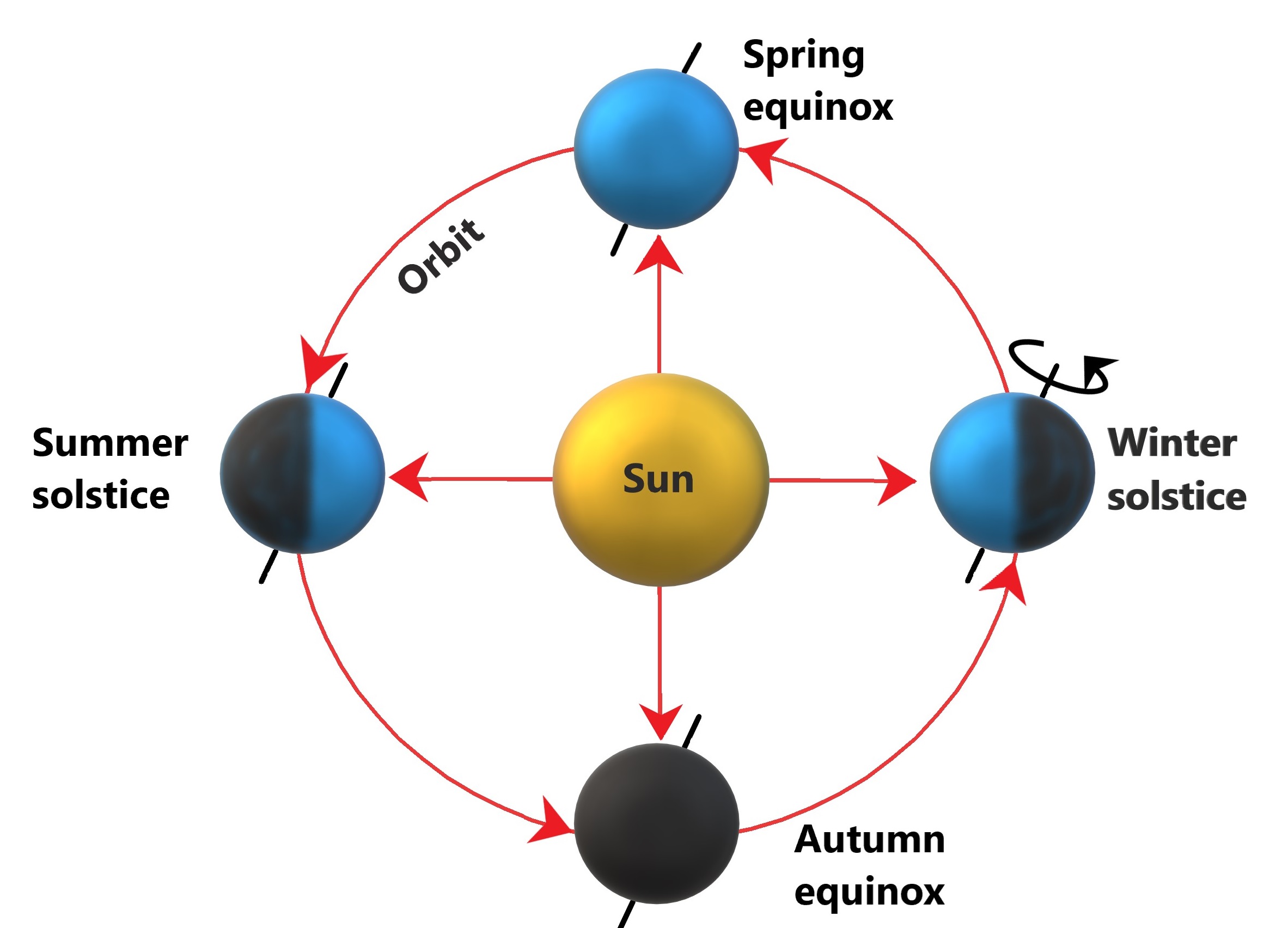} 
\caption{ The schematic Earth orbit about the Sun\cite{google18} (shown not to the real scale of the Sun and Earth) with the indication of the solar irradiance at different phases of the orbit \cite{Eddy1976, Eddy1983}. The arrows coming from the centre of the Sun in two perpendicular directions are symbolic axis of the Earth orbital motion: vertical one is shorter and the horizontal one is slightly longer according to the Earth orbit eccentricity. The other arch arrows are also symbolic showing the direction of the Earth rotation about the Sun (anti-clockwise). The Earth axis is shown by the thin lines coming from the North and South poles, the dark parts of the Earth disk show the night, and the blue ones show the day. The Earth latitudes are shown by the light lines on the disk, while the current angle of the Earth axis inclination from the perpendicular to the ecliptics is shown in the winter solstice (the right disk). }  
\label{earth_orbit}
\end{figure*}

 The increase of the solar irradiance at these times is expected to lead to the increase of the terrestrial temperature\cite{Eddy1976, Shirley1990} in the Northern hemisphere where the most  solar observatories measuring the terrestrial temperature are located. Since Akasofu\cite{akasofu10} derived the rate of the temperature increase in the past centuries to be about 0.5C per 100 years (see Fig. \ref{irradiance}, bottom plot). Therefore, with a very conservative extrapolation of this temperature into the next 5.5 centuries, following the parabola of the baseline wave caused by SIM, we expect an increase of the terrestrial temperature in the Northern hemisphere from the current magnitude by about 2.5 C or slightly higher. This increase is caused solely by the Sun's rotation about the barycenter of the solar system as it is shown in Fig.\ref{irradiance}, top plot. Given the fact that these temperature variations have already happened on the Earth many thousand times in the past, one expects the Earth-Sun system to handle this increase in its usual ways. Of course, any human-induced contributions can make this increase  more unpredictable and difficult to handle if they will override the effects on the temperature induced by the Sun.
 
We have to emphasise that there still will be, of course, the usual magnetic field and temperature oscillations caused by standard solar activity cycles of 11 and 350-400 years as reported before\cite{zhar15} occurring on top of these baseline oscillations caused by SIM. As result, the solar irradiance and terrestrial temperature are expected to oscillate around this baseline for the next 700 years while increasing during the maxima of 11 year and 350 year solar cycles and decreasing during their minima, similarly to the natural temperature variations oscillating about the temperature baseline  shown by black line in the plot by Akasofu\cite{akasofu10}(see Fig. \ref{irradiance}, bottom plot and Akasofu's Fig. 9\cite{akasofu10}). However, during the next two grand solar minima, which are expected to occur in 2020-2055 (Modern grand solar minimum lasting for 3 solar cycles) and  in 2370-2415 (future grand solar minimum lasting for 4 cycles) (see Fig. 3 in Zharkova et al.\cite{zhar15}) a decrease of the terrestrial temperature is expected to be similar to those during the Maunder Minimum and, definitely, substantially larger than natural temperature fluctuations shown in the Akasofu's plot\cite{akasofu10, lean1995}. Note, these oscillations of the estimated terrestrial temperature  do not include any human-induced factors, but only the effects of solar activity itself and solar inertial motion.

 \section*{Conclusions}
  
Until recently, solar activity was accepted to be one of the important factors defining the temperature on Earth and other planets. In this paper we reproduced the summary curve of the solar magnetic field associated with solar activity\cite{shepherd14, zhar15} for the one hundred thousand years  backward by using the formulas describing the sum of the two principal components found from the full disk solar magnetograms. In the past 3000 years the summary curve  shows the solar activity for every 11 years and occurrence of 9 grand solar cycles of 350-400 years, which are  caused by the  beating effects of  two waves generated at the inner and outer layers inside the solar interior with close but not equal frequencies\cite{zhar15}. 

The resulting summary curve reveals a remarkable resemblance to the sunspot and terrestrial activity reported in the past millennia including the significant grand solar minima: Maunder Minimum (1645-1715),  Wolf  minimum (1200),  Oort minimum (1010-1050),  Homer  minimum (800-900 BC) combined with the grand solar maxima: the medieval warm  period (900-1200), the Roman warm  period (400-10BC) etc. It also predicts the upcoming  grand  solar minimum, similar to Maunder Minimum, which starts in 2020 and will last until 2055.   

A reconstruction of solar total irradiance suggests that there is an increase in the cycle-averaged total solar irradiance (TSI) since  the Maunder minimum  by a value of about $1–1.5$ $Wm^{-2}$\cite{krivova11}. This increase is closely correlated with the similar  increase of the  average terrestrial temperature\cite{Shirley1990, akasofu10}. Moreover,  from  the summary curve for the past 100 thousand years we found the similar oscillations of the baseline of  magnetic field with a period of $1950\pm95$ years (a super-grand  solar cycle)  by filtering out the large-scale oscillations in 11 year cycles.  The last minimum of a super-grand cycle occurred at the beginning of Maunder minimum. Currently, the baseline  magnetic field (and solar irradiance) are increasing to reach its maximum at{\color{blue}  $~$2700}, after which  the baseline magnetic field become decreasing for another 1000 years.  

The  oscillations of the baseline of solar magnetic field are likely to be caused by the solar inertial  motion about the barycentre of the solar system caused by large planets. This, in turn, is closely linked to an increase of solar irradiance caused by the positions of the Sun either closer to aphelion and autumn equinox or perihelion and spring equinox. Therefore,  the oscillations of the baseline define the global trend of solar magnetic field and solar irradiance over a period of about 2100 years. In the current millennium since Maunder minimum we have the increase of the baseline magnetic field and solar irradiance for another 700 years. This increase would lead to the terrestrial temperature  increase as noted by Akasofu\cite{akasofu10} during the past two hundred years.  Based on the growth rate of 0.5C per 100 years\cite{akasofu10} for the  terrestrial temperature since Maunder minimum, one can anticipate that the increase of the solar baseline magnetic field expected to occur up to {\color{blue} 2700}  because of SIM will lead, in turn,  to  the  increase of the terrestrial baseline temperature since MM by 1.3$^\circ$C (in 2100) and, at least, by  2.5-3$^\circ$C (in {\color{blue} 2700}).

Naturally, on top of this increase of the baseline terrestrial  temperature, there are  imposed much larger temperature oscillations caused by standard solar activity cycles of 11 and 350-400 years and terrestrial causes. The terrestrial temperature is expected to grow during maxima of 11 year solar cycles and to decrease during their minima. Furthermore, the  substantial temperature  decreases are  expected during the two grand minima\cite{lean1995} to occur in 2020-2055 and  2370-2415\cite{zhar15}, whose magnitudes cannot be yet predicted and need further investigation. These oscillations of the estimated terrestrial temperature  do not include any human-induced factors, which were outside the scope of the current paper.
\clearpage

\section*{Acknowledgments}

The authors wish to thank the Wilcox Solar Observatory staff for providing the magnetic data for the whole disk Sun since 1976, which were used in the current research.  VZ and SZ appreciate the support of the AFOSR grant funded by the US Airforce. EP wish to thank Russian Science Foundation (Project 17-11-01052).  EP wishes  to thank Northumbria University for their kind hospitality and warm reception during her visit funded by the Royal Society when the paper was initiated.  

%\clearpage
%\bibliography{zhark_nature2}

\begin{thebibliography}{10}
\expandafter\ifx\csname url\endcsname\relax
  \def\url#1{\texttt{#1}}\fi
\expandafter\ifx\csname urlprefix\endcsname\relax\def\urlprefix{URL }\fi
\providecommand{\bibinfo}[2]{#2}
\providecommand{\eprint}[2][]{\url{#2}}

\bibitem{pesnell08}
\bibinfo{author}{{Pesnell}, W.~D.}
\newblock \bibinfo{title}{{Predictions of Solar Cycle 24}}.
\newblock \emph{\bibinfo{journal}{\solphys}} \textbf{\bibinfo{volume}{252}},
  \bibinfo{pages}{209--220} (\bibinfo{year}{2008}).


\bibitem{zhar08}
\bibinfo{author}{{Zharkov}, S.}~I., \bibinfo{author}{{Gavrjuseva}, E~V.} \&
  \bibinfo{author}{{Zharkova}, V.~V.}
\newblock \bibinfo{title}{{The Observed Long- and Short-Term Phase Relation between the Toroidal and Poloidal Magnetic Fields in Cycle 23}} 
newblock \emph{\bibinfo{journal}{\solphys}} \textbf{\bibinfo{volume}{248}},
\bibinfo{pages}{339-358} (\bibinfo{year}{2008}).
 
 \bibitem{stix76}
\bibinfo{author}{{Stix}, M.}, 
\newblock \bibinfo{title}{{Differential rotation and the solar dynamo}}.
\newblock \emph{\bibinfo{journal}{Astronomy and Astrophysics}}
  \textbf{\bibinfo{volume}{47}}, \bibinfo{pages}{243-254}
   (\bibinfo{year}{1976}). 
   
\bibitem{zharkova12}
\bibinfo{author}{{Zharkova}, V.~V.}, \bibinfo{author}{{Shepherd}, S.~J.} \&
  \bibinfo{author}{{Zharkov}, S.~I.}
\newblock \bibinfo{title}{{Principal component analysis of background and
  sunspot magnetic field variations during solar cycles 21-23}}.
\newblock \emph{\bibinfo{journal}{\mnras}} \textbf{\bibinfo{volume}{424}},
  \bibinfo{pages}{2943--2953} (\bibinfo{year}{2012}).

\bibitem{shepherd14}
\bibinfo{author}{{Shepherd}, S.~J.}, \bibinfo{author}{{Zharkov}, S.~I.} \&
  \bibinfo{author}{{Zharkova}, V.~V.}
\newblock \bibinfo{title}{{Prediction of Solar Activity from Solar Background
  Magnetic Field Variations in Cycles 21-23}}.
\newblock \emph{\bibinfo{journal}{\apj}} \textbf{\bibinfo{volume}{795}},
  \bibinfo{pages}{46} (\bibinfo{year}{2014}).

\bibitem{zhar15}
\bibinfo{author}{{Zharkova}, V.~V.}, \bibinfo{author}{{Shepherd}, S.~J.},
  \bibinfo{author}{{Popova}, E.} \& \bibinfo{author}{{Zharkov}, S.~I.}
\newblock \bibinfo{title}{{Heartbeat of the Sun from Principal Component
  Analysis and prediction of solar activity on a millenium timescale}}.
\newblock \emph{\bibinfo{journal}{Nature Scientific Reports}}
  \textbf{\bibinfo{volume}{5}}, \bibinfo{pages}{15689} (\bibinfo{year}{2015}).

\bibitem{popova13}
\bibinfo{author}{{Popova}, E.}, \bibinfo{author}{{Zharkova}, V.} \&
  \bibinfo{author}{{Zharkov}, S.}
\newblock \bibinfo{title}{{Probing latitudinal variations of the solar magnetic
  field in cycles 21-23 by Parker's Two-Layer Dynamo Model with meridional
  circulation}}.
\newblock \emph{\bibinfo{journal}{Annales Geophysicae}}
  \textbf{\bibinfo{volume}{31}}, \bibinfo{pages}{2023--2038}
  (\bibinfo{year}{2013}).

\bibitem{zhao2013}
\bibinfo{author}{{Zhao}, J.}, \bibinfo{author}{{Bogart}, R.~S.},
  \bibinfo{author}{{Kosovichev}, A.~G.}, \bibinfo{author}{{Duvall}, T.~L., Jr.}
  \& \bibinfo{author}{{Hartlep}, T.}
\newblock \bibinfo{title}{{Detection of Equatorward Meridional Flow and
  Evidence of Double-cell Meridional Circulation inside the Sun}}.
\newblock \emph{\bibinfo{journal}{\apjl}} \textbf{\bibinfo{volume}{774}},
  \bibinfo{pages}{L29} (\bibinfo{year}{2013}).
\newblock \eprint{1307.8422}.

\bibitem{zhar05}
\bibinfo{author}{{Zharkov}, S.}~I., \bibinfo{author}{{Zharkova}, V~V.} \&
  \bibinfo{author}{{Ipson}, S.~S.}
\newblock \bibinfo{title}{{Statistical Properties Of Sunspots In 1996 2004: I. Detection, North South Asymmetry And Area Distribution}} 
newblock \emph{\bibinfo{journal}{\solphys}} \textbf{\bibinfo{volume}{228}},
\bibinfo{pages}{337-357} (\bibinfo{year}{2005}).
 
\bibitem{temmer02}
\bibinfo{author}{{Temmer}, M.}~I., \bibinfo{author}{{Veronig}, A.} \&
  \bibinfo{author}{{Hanslmeier}, A.}
\newblock \bibinfo{title}{{Hemispheric Sunspot Numbers R$_{n}$ and R$_{s}$: Catalogue and N-S asymmetry analysis}} 
newblock \emph{\bibinfo{journal}{\aap}} \textbf{\bibinfo{volume}{490}},
\bibinfo{pages}{707-715} (\bibinfo{year}{2002}).

\bibitem{belucz_dikpati2013}
\bibinfo{author}{{Belucz}, B.} \& \bibinfo{author}{{Dikpati}, M.} 
\newblock \bibinfo{title}{{Role of Asymmetric Meridional Circulation in Producing North-South Asymmetry in a Solar Cycle Dynamo Model}} 
newblock \emph{\bibinfo{journal}{\apj}} \textbf{\bibinfo{volume}{779}},
\bibinfo{pages}{4-10} (\bibinfo{year}{2013}).
 
\bibitem{shetye_dikpati2015}
\bibinfo{author}{{Shetye}, J. and {Tripathi}, D. and {Dikpati}, M.}
\newblock \bibinfo{title}{{Observations and Modeling of North-South Asymmetries Using a Flux Transport Dynamo}} 
newblock \emph{\bibinfo{journal}{\apj}} \textbf{\bibinfo{volume}{799}},
\bibinfo{pages}{220-230} (\bibinfo{year}{2015}).

\bibitem{Parker55}
\bibinfo{author}{{Parker}, E.~N.}
\newblock \bibinfo{title}{{Hydromagnetic Dynamo Models.}}
\newblock \emph{\bibinfo{journal}{\apj}} \textbf{\bibinfo{volume}{122}},
  \bibinfo{pages}{293} (\bibinfo{year}{1955}).

\bibitem{axel2005}
\bibinfo{author}{{Brandenburg}, A.} \& \bibinfo{author}{{Subramanian}, K.}
\newblock \bibinfo{title}{{Astrophysical magnetic fields and nonlinear dynamo
  theory}}.
\newblock \emph{\bibinfo{journal}{\physrep}} \textbf{\bibinfo{volume}{417}},
  \bibinfo{pages}{1--209} (\bibinfo{year}{2005}).
%\newblock \eprint{astro-ph/0405052}.

\bibitem{jones10}
\bibinfo{author}{{Jones}, C.~A.}, \bibinfo{author}{{Thompson}, M.~J.} \&
  \bibinfo{author}{{Tobias}, S.~M.}
\newblock \bibinfo{title}{{The Solar Dynamo}}.
\newblock \emph{\bibinfo{journal}{\ssr}} \textbf{\bibinfo{volume}{152}},
  \bibinfo{pages}{591--616} (\bibinfo{year}{2010}).
  
 \bibitem{popova17}
\bibinfo{author}{{Popova}, E.}, \bibinfo{author}{{Zharkova}, V.}, \bibinfo{author}{{Shepherd}, S.~J.} \&
  \bibinfo{author}{{Zharkov}, S.}
\newblock \bibinfo{title}{{On a role of quadruple component of magnetic field in defining solar activity in grand cycles}}.
\newblock \emph{\bibinfo{journal}{Journal of Atmospheric and Solar-Terrestrial Physics}}
  \textbf{\bibinfo{volume}{176}}, \bibinfo{pages}{61--71}
  (\bibinfo{year}{2018}).

\bibitem{2012ApJ...761L..13K}
\bibinfo{author}{{Karak}, B.~B.} \& \bibinfo{author}{{Nandy}, D.}
\newblock \bibinfo{title}{{Turbulent Pumping of Magnetic Flux Reduces Solar
  Cycle Memory and thus Impacts Predictability of the Sun's Activity}}.
\newblock \emph{\bibinfo{journal}{\apjl}} \textbf{\bibinfo{volume}{761}},
  \bibinfo{pages}{L13} (\bibinfo{year}{2012}).
%\newblock \eprint{1206.2106}.

\bibitem{zhar17}
\bibinfo{author}{{Zharkova}, V.~V.}, \bibinfo{author}{{Shepherd}, S.~J.},
  \bibinfo{author}{{Popova}, E.} \& \bibinfo{author}{{Zharkov}, S.~I.}
\newblock \bibinfo{title}{{Reply to comment by Usoskin (2017) on the paper "On a role of quadruple component of magnetic field in defining solar activity in grand cycles}}.
\newblock \emph{\bibinfo{journal}{Journal of Atmospheric and Solar-Terrestrial Physicss}}
  \textbf{\bibinfo{volume}{176}}, \bibinfo{pages}{72-82}
   (\bibinfo{year}{2018}).
   
\bibitem{zhar18}
\bibinfo{author}{{Zharkova}, V.~V.}, \bibinfo{author}{{Shepherd}, S.~J.},
  \bibinfo{author}{{Popova}, E.} \& \bibinfo{author}{{Zharkov}, S.~I.}
\newblock \bibinfo{title}{{Reinforcing a Double Dynamo Model with Solar-Terrestrial Activity in the Past Three Millennia}}.
\newblock \emph{\bibinfo{journal}{Proc. IAU Symposium}}
  \textbf{\bibinfo{volume}{335}}, \bibinfo{pages}{211-215}
   (\bibinfo{year}{2018}).
   
\bibitem{arnold_libby1949}
\bibinfo{author}{{Arnold}, J.~R.} \& \bibinfo{author}{{Libby}, W.~F.}
\newblock \bibinfo{title}{{Age Determinations by Radiocarbon Content: Checks
  with Samples of Known Age}}.
\newblock \emph{\bibinfo{journal}{Science}} \textbf{\bibinfo{volume}{110}},
  \bibinfo{pages}{678--680} (\bibinfo{year}{1949}).
  
\bibitem{baade1934}
\bibinfo{author}{{Baade}, W.} \& \bibinfo{author}{{Zwicky}, F.}
\newblock \bibinfo{title}{{Cosmic Rays from Super-novae}}.
\newblock \emph{\bibinfo{journal}{Proceedings of the National Academy of
  Science}} \textbf{\bibinfo{volume}{20}}, \bibinfo{pages}{259--263}
  (\bibinfo{year}{1934}).

\bibitem{hays76}
\bibinfo{author}{{Hays}, J.~D.}, \bibinfo{author}{{Imbrie}, J.}\&
\bibinfo{author}{{Shackelton}, N.~J.}
\newblock \emph{\bibinfo{title}{{Variations in the Earth's Orbit: Pacemaker of the Ice Ages}}}.
\newblock \emph{\bibinfo{journal}{science}} \textbf{\bibinfo{volume}{194}},
  \bibinfo{pages}{1121--1126}
  (\bibinfo{year}{1976}).
 
\bibitem{milankovich98}
\bibinfo{author}{{Milankovich}, M.},  
\newblock \emph{\bibinfo{title}{{Canon of Insolation and the Ice Age Problem}}}.
\newblock \emph{\bibinfo{book}{Belgrade: Zavod za Udzbenike i Nastavna Sredstva}}, {ISBN 86-17-06619-9 }  (\bibinfo{year}{1998}).  

\bibitem{abe13}
\bibinfo{author}{{Abe-Ouchi}, A.}, \bibinfo{author}{{Saito}, F.}, \bibinfo{author}{{Kamamura}, K.}, \bibinfo{author}{{Raymo}, M.~E.}, \bibinfo{author}{{Okuno}, J.}, \bibinfo{author}{{Takashi}, K.}\&
\bibinfo{author}{{Blatter}, H.}
\newblock \emph{\bibinfo{title}{{Insolation-driven 100,000-year glacial cycles and hysteresis of ice-sheet volume}}}.
\newblock \emph{\bibinfo{journal}{nature}} \textbf{\bibinfo{volume}{500}},
  \bibinfo{pages}{7461--}
  (\bibinfo{year}{2013}).

 \bibitem{rial03}
\bibinfo{author}{{Rial}, J.~A.}  
\newblock \emph{\bibinfo{title}{{Earth's orbital Eccentricity and the rhythm of the Pleistocene ice ages: the concealed pacemake}}}.
\newblock \emph{\bibinfo{journal}{Global and Planetary Change}} \textbf{\bibinfo{volume}{41}},
  \bibinfo{pages}{81--93}
  (\bibinfo{year}{2003}).
   
  \bibitem{akasofu10}
\bibinfo{author}{{Akasofu}, P.} 
\newblock \bibinfo{title}{{On the recovery from the Little Ice Age}}.
\newblock \emph{\bibinfo{journal}{Natural Science}} \textbf{\bibinfo{volume}{2}},
  \bibinfo{pages}{1211-1224} (\bibinfo{year}{2010}).
           
\bibitem{krivova11}
\bibinfo{author}{{Krivova}, N.~A.} \& \bibinfo{author}{{Solanki}, S.~K.}
\newblock \bibinfo{title}{{Towards a long-term record of solar total and spectral irradiance}}.
\newblock \emph{\bibinfo{journal}{Journal of Atmopsheric and Solar-Terrestrial Physics}} \textbf{\bibinfo{volume}{73}},
  \bibinfo{pages}{223-234} (\bibinfo{year}{2011}).
  
 {\color{blue} 
 \bibitem{solanki04}
\bibinfo{author}{{Solanki}, S.~K.}, \bibinfo{author}{{Usoskin}, I.~G.},
\bibinfo{author}{{Kromer}, B.~K.}, \bibinfo{author}{{M. Schüssler}, M.} and \bibinfo{author}{{Beer}, J.}
\newblock \bibinfo{title}{{Unusual activity of the Sun during recent decades compared to the previous 11,000 years}}.
\newblock \emph{\bibinfo{journal}{\nat}} \textbf{\bibinfo{volume}{431}},
  \bibinfo{pages}{91084–1087} (\bibinfo{year}{2004}).}
  
 \bibitem{Fairbridge1987}
\bibinfo{author}{{Fairbridge}, R.~W.} \& \bibinfo{author}{{Shirley}, J.~H.}
\newblock \bibinfo{title}{{Prolonged minima and the 179-yr cycle of the solar
  inertial motion}}.
\newblock \emph{\bibinfo{journal}{\solphys}} \textbf{\bibinfo{volume}{110}},
  \bibinfo{pages}{191--210} (\bibinfo{year}{1987}).
  
 {\color{blue}  
\bibitem{Steinhilber12}
\bibinfo{author}{{Steinhilber}, F.}, \bibinfo{author}{{Abreu}, J.A.}, \bibinfo{author}{{Beer}, J.},  \bibinfo{author}{{Brunnera}, I.}, \bibinfo{author}{{Christl}, M.}, 
\bibinfo{author}{{Fischer}, H.}, \bibinfo{author}{{Heikkiläd}, U.},  \bibinfo{author}{{Kubik}, P. W.}, \bibinfo{author}{{Manna}, M.},  \bibinfo{author}{{McCracken}, K.G.}, \bibinfo{author}{{Miller}, H},  \bibinfo{author}{{Miyahara}, H.}, \bibinfo{author}{{Oerter}, H.}, \&  \bibinfo{author}{{Wilhelms}, F.},
\newblock \bibinfo{title}{{9,400 years of cosmic radiation and solar activity from ice cores and tree rings}}.
\newblock \emph{\bibinfo{journal}{Proc. of the National Academy of Sciences}} \textbf{\bibinfo{volume}{1096}},
  \bibinfo{pages}{5967-5971} (\bibinfo{year}{2012}).
\newblock \eprint{1202.3554}.}
  
  \bibitem{scafetta2014}
\bibinfo{author}{{Scafetta}, N.}
\newblock \bibinfo{title}{{Discussion on the spectral coherence between
  planetary, solar and climate oscillations: a reply to some critiques}}.
\newblock \emph{\bibinfo{journal}{Astrophysics and Space Science}}
  \textbf{\bibinfo{volume}{354}}, \bibinfo{pages}{275--299}
  (\bibinfo{year}{2014}).
\newblock \eprint{1412.0250}.

\bibitem{obridko2014}
\bibinfo{author}{{Obridko}, V.} \& \bibinfo{author}{{Nagovitsyn}, Y.}
\newblock \bibinfo{title}{{Solar activity over different timescales}}.
\newblock In \emph{\bibinfo{booktitle}{40th COSPAR Scientific Assembly}},
  vol.~\bibinfo{volume}{40} of \emph{\bibinfo{series}{COSPAR Meeting}}
  (\bibinfo{year}{2014}).
 
\bibitem{vieira2011}
\bibinfo{author}{{Vieira}, L-I.~A.}, \bibinfo{author}{{Solanki}, S.~K.},  \bibinfo{author}{{Krivova}, N.~A.} \& \bibinfo{author}{{Usoskin}, I.} 
\newblock \bibinfo{title}{{Evolution of the solar irradiance during the Holocene}}.
\newblock \emph{\bibinfo{journal}{\aa}}
  \textbf{\bibinfo{volume}{531}}, \bibinfo{pages}{A6}
  (\bibinfo{year}{2011}).
    
\bibitem{Kuklin1976}
\bibinfo{author}{{Kuklin}, G.~V.}
\newblock \bibinfo{title}{{Cyclical and Secular Variations of Solar Activity}}.
\newblock \emph{\bibinfo{Proceedings}{Basic Mechanisms of Solar Activity, Proceedings from IAU Symposium  Prague, Czechoslovakia, 25-29 August 1975}}
  \textbf{\bibinfo{volume}{71}}, \bibinfo{pages}{147} (\bibinfo{year}{1976}).

\bibitem{charvatova1988}
\bibinfo{author}{{Charvatova}, I.} \
\newblock \bibinfo{title}{{The solar motion and the variability of solar activity}}.
\newblock \emph{\bibinfo{journal}{\jasr}} \textbf{\bibinfo{volume}{8}},
  \bibinfo{pages}{147--150} (\bibinfo{year}{1988}).

\bibitem{Shirley1990}
\bibinfo{author}{{Shirley}, J.~H.} \& \bibinfo{author}{{Sperber}, K.~R.} \& \bibinfo{author}{{Fairbridge}, R.~W.}
\newblock \bibinfo{title}{{Sun's inertial motion and luminosity}}.
\newblock \emph{\bibinfo{journal}{\solphys}} \textbf{\bibinfo{volume}{127}},
  \bibinfo{pages}{379--392} (\bibinfo{year}{1990}).

\bibitem{palus2007}
\bibinfo{author}{{Palu{\v s}}, M. and {Kurths}, J. and {Schwarz}, U. and {Seehafer}, N. and 
	{Novotn{\'a}}, D. and {Charv{\'a}tov{\'a}}, I.},
\newblock \bibinfo{title}{{The solar activity cycle is weakly synchronized with the solar inertial motion}}.
\newblock \emph{\bibinfo{journal}{Physics Letters A}} \textbf{\bibinfo{volume}{365}},
  \bibinfo{pages}{421--428} (\bibinfo{year}{2007}).
  
\bibitem{mackey07}
\bibinfo{author}{{Mackey}, R.} 
\newblock \bibinfo{title}{{Rhodes Fairbridge and the idea that the solar system regulates the Earth’s climate}},
\newblock \emph{\bibinfo{journal}{Journal of Coastal Research, (Proceedings of the Ninth International Coastal Symposium, Gold Coast, Australia)}}, \textbf{\bibinfo{volume}{SI 50}},
  \bibinfo{pages}{955--968} (\bibinfo{year}{2007}). 
  
\bibitem{charvatova2000}
\bibinfo{author}{{Charvatova}, I.} \
\newblock \bibinfo{title}{{Can origin of the 2400-year cycle of solar activity be caused by solar inertial motion?}}.
\newblock \emph{\bibinfo{journal}{AnnGeo}} \textbf{\bibinfo{volume}{18}},
  \bibinfo{pages}{399--405} (\bibinfo{year}{2000}).

\bibitem{zaqarashvili97}
\bibinfo{author}{{Zaqarashvili}, T.~V.} 
\newblock \bibinfo{title}{{On a Possible Generation Mechanism for the Solar Cycle}}.
\newblock \emph{\bibinfo{journal}{\apj}} \textbf{\bibinfo{volume}{487}},
  \bibinfo{pages}{930-935} (\bibinfo{year}{1997}).

\bibitem{Abreu2012}
\bibinfo{author}{{Abreu}, J.~A.} \& \bibinfo{author}{{Beer}, J.} \& \bibinfo{author}{{Ferriz-Mas}, A.} \& \bibinfo{author}{{McCracken}, K.~G.} \& \bibinfo{author}{{Steinhilber}, F.}
\newblock \bibinfo{title}{{Is there a planetary influence on solar activity?}}.
\newblock \emph{\bibinfo{journal}{\aa}} \textbf{\bibinfo{volume}{548}},
  \bibinfo{pages}{A88} (\bibinfo{year}{2012}).

\bibitem{Goldreich89}
\bibinfo{author}{{Goldreich}, P.} \& \bibinfo{author}{{Nicholson}, P.~D.}
\newblock \bibinfo{title}{{Tidal friction in early-type stars}}.
\newblock \emph{\bibinfo{journal}{\apj}} \textbf{\bibinfo{volume}{342}},
  \bibinfo{pages}{1079-1084} (\bibinfo{year}{1989}).
  
\bibitem{Barker10}
\bibinfo{author}{{Barker}, A.~J.} \& \bibinfo{author}{{Ogilvie}, G.~I.}
\newblock \bibinfo{title}{{On internal wave breaking and tidal dissipation near the centre of a solar-type star}}.
\newblock \emph{\bibinfo{journal}{\mnras}} \textbf{\bibinfo{volume}{404}},
  \bibinfo{pages}{1849-1864} (\bibinfo{year}{2010}).
  
\bibitem{Charbonneau1999}
\bibinfo{author}{{Charbonneau}, P. and {Christensen-Dalsgaard}, J. and {Henning}, R. and 
	{Larsen}, R.~M. and {Schou}, J. and {Thompson}, M.~J. and {Tomczyk}, S.},
\newblock \bibinfo{title}{{Helioseismic Constraints on the Structure of the Solar Tachocline}}.
\newblock \emph{\bibinfo{journal}{\apj}} \textbf{\bibinfo{volume}{527}},
  \bibinfo{pages}{445-460} (\bibinfo{year}{1999}).
  
  
\bibitem{Cameron13}
\bibinfo{author}{{Cameron}, R.~H.} \& \bibinfo{author}{{Schussler}, M.}
\newblock \bibinfo{title}{{No evidence for planetary evidence on solar activity}}.
\newblock \emph{\bibinfo{journal}{\aa}} \textbf{\bibinfo{volume}{557}},
  \bibinfo{pages}{A83} (\bibinfo{year}{2013}).
 
 {\color{blue} 
\bibitem{folkner2014}
\bibinfo{author}{{Folkner}, W.~M.}, \bibinfo{author}{{Williams}, J.~G.},
  \bibinfo{author}{{Boggs}, D.~H.}, \bibinfo{author}{{Park}, R.~S.} \&
  \bibinfo{author}{{Kuchynka}, P.}
\newblock \bibinfo{title}{{The Planetary and Lunar Ephemerides DE430 and
  DE431}}.
\newblock \emph{\bibinfo{journal}{The Interplanetary Network Progress Report)}}
  \textbf{\bibinfo{volume}{42}}, \bibinfo{pages}{1--10} (\bibinfo{year}{2014}).}
  
\bibitem{Eddy1976}
\bibinfo{author}{{Eddy}, J.~A.}
\newblock \bibinfo{title}{{The Maunder Minimum}}.
\newblock \emph{\bibinfo{journal}{Science}} \textbf{\bibinfo{volume}{192}},
  \bibinfo{pages}{1189--1202} (\bibinfo{year}{1976}).
 
\bibitem{Eddy1983}
\bibinfo{author}{{Eddy}, J.~A.}
\newblock \bibinfo{title}{{The Maunder Minimum - A reappraisal}}.
\newblock \emph{\bibinfo{journal}{\solphys}} \textbf{\bibinfo{volume}{89}},
  \bibinfo{pages}{195--207} (\bibinfo{year}{1983}).

 \bibitem{dikpati2006}
\bibinfo{author}{{Dikpati}, M. and {Gilman}, P.~A. and {Dikpati}, M.}
\newblock \bibinfo{title}{{Penetration of dynamo-generated magnetic fields into the sun's radiative interior}}, 
newblock \emph{\bibinfo{journal}{\apj}} \textbf{\bibinfo{volume}{638}},
\bibinfo{pages}{564-575} (\bibinfo{year}{2006}).

{\color{blue}  
\bibitem{lean1995}
\bibinfo{author}{{Lean}, J.}, \bibinfo{author}{{Beer}, J.} \&
  \bibinfo{author}{{Bradley}, R.}
\newblock \bibinfo{title}{{Reconstruction of Solar Irradiance Since 1610:
  Implications for Climate Change}}.
\newblock \emph{\bibinfo{journal}{Geophysical Research Letters}}
  \textbf{\bibinfo{volume}{22}}, \bibinfo{pages}{3195--3198}
  (\bibinfo{year}{1995}).}
  
\end{thebibliography}

\section*{Author contributions statement}

 VZ, S.S. and S.Z. analysed the data with PCA, S.S., V.Z. and S.Z. conducted the data prediction with Euriqa for the summary curves, E.P. developed the dynamo models and helped to build the plots for observations.  All authors analysed the results and reviewed the manuscript. 

\section*{Additional information}

\noindent The authors declare no competing financial or non-financial interests.  

\noindent The datasets generated during and/or analysed during the current study are available from the corresponding author on reasonable request.

\end{document}